Paper follows in LaTeX (AIP RevTeX version 2.1 macros) format.
Also, the EPSF macros for Encapsulated PostScript figures
is used, in conjunction with the "dvips" converter by
Tomas Rokicki <rokicki@neon.stanford.edu>.
PostScript figures are not included.  The entire preprint including
figures can be made available as a compressed PostScript file (400 kB)
via anonymous FTP, if people send E-mail to: Ole.H.Nielsen@uni-c.dk.

\documentstyle[preprint,revtex,eqsecnum]{aps}

\input epsf	
\draft
\tightenlines

\begin{document}

\begin{title}
Melting a copper cluster: Critical droplet theory
\end{title}

\author{Ole H. Nielsen}
\begin{instit}
UNI$\bullet$C,
Building 305, Technical University of Denmark, DK-2800 Lyngby, Denmark
\end{instit}

\author{James P. Sethna\cite{Sethna-addr},
Per Stoltze, Karsten W. Jacobsen and Jens K. N{\o}rskov}
\begin{instit}
Laboratory of Applied Physics,
Building 307, Technical University of Denmark,\\ DK-2800 Lyngby, Denmark
\end{instit}

\begin{abstract}

We simulate the melting of a 71~{\AA} diameter cluster of Cu.  At low
temperatures the crystal exhibits facets.  With increasing temperatures
the open facets pre--melt, the melted regions coalesce into a liquid
envelope containing a crystalline nucleus, and the nucleus finally goes
unstable to the supercooled liquid.  Using critical droplet theory and
experimental data for Cu, we explain the thermodynamics of the
coexistence region.  The width of the transition scales as (Number of
particles)$^{-1/4}$.

\end{abstract}
\receipt{July 10, 1992}

We simulate the melting of an approximately spherical, 71~{\AA}
diameter, 16727 atom cluster of Cu, using molecular dynamics with
effective medium theory interactions.\cite{Norskov,Stoltze}  By using a
realistic interaction potential we are not only able to make specific
experimental predictions, but we are also able to compare our results
to analytical calculations using experimentally determined latent
heats, specific heats, surface tensions, etc.  We analyze our results
in three contexts.  First, we observe many of the effects predicted in
the theory of equilibrium crystal shapes,\cite{Wortis} including liquid
regions on open facets which can either be interpreted as
pre--melting\cite{Premelting} or ``liquid
lenses''.\cite{Lowen,Nozieres} Second, we use critical droplet
theory\cite{Langer} to quantitatively explain the thermodynamics of the
transition.  By working at fixed energy (rather than fixed temperature)
the first--order transition unfolds, revealing an interesting partially
melted region in which adding more energy to the cluster {\it reduces}
its mean temperature (the system has a negative specific heat).
Third, we make contact to the formal theory of finite--size effects in
first--order phase transitions.  We show that the broadening we observe
in the transition scales as the number of atoms $N^{-1/4}$, which for
large $N$ dominates the other studied broadening
mechanisms.\cite{Privman}

The molecular dynamics simulations of the properties of this Cu cluster
are carried out using the Verlet algorithm for time-integration of the
equation of motion.  The simulations employ Andersen
thermalization\cite{HCA} for obtaining a given temperature or total
energy of the system.  The calculations were done on a massively
parallel Connection Machine CM-200 computer, using an algorithm where
the face-centered cubic computational box was subdivided into a $16^3$
grid.  A computational processor is assigned to each grid point
computing for the atoms in its associated grid sub-box.  The
interatomic interactions of the effective--medium theory (which extend
roughly to fourth neighbors) are carried out by regular communication
of atomic properties inside $5^3$ sub-boxes surrounding each atom,
assuring that all non-zero interactions are accounted for correctly, as
in the serial-computer algorithm in ref. \cite{Stoltze}.  These
algorithms will be described in a succeeding paper.\cite{OHN:MD}

In the present study of the solid-liquid phase transition, it is
natural to control the total energy parameter and measure the
temperature as the average kinetic energy of the atoms.  The
fluctuation $\langle (\Delta T)^2 \rangle$ of the equilibrium
temperature $T$ at constant energy can be shown to be
$k_B T^2 [ 2 / (3 N k_B) - 1 / C ]$,
where $N$ is the number of atoms, and $C$ denotes the specific heat of
the cluster as a whole.  The measured fluctuations agree well with this
formula except near the instability point where the critical slowing
down (see below) makes sufficient statistics impractical to collect.

The Cu cluster was initially a spherically truncated fcc crystal,
heated to 1282~K in order to melt about 3 layers of surface atoms.
This cluster was both cooled and heated, respectively, at a
sufficiently slow rate to maintain thermal equilibrium.  Figure~1 shows
the facet distribution developing after the cooling.  The relative
areas of these facets in equilibrium depend on their relative surface
energies through the Wulff construction,\cite{Wortis} and are therefore
quite sensitive to the model of interatomic interactions.  The
effective--medium theory has been shown to describe the various Cu
surface energies with good accuracy.\cite{Cu-surfaces}

\bigskip
{\bf Figure 1.}
The faceted copper cluster at low temperature (522~K), with prominent
(111) and (100) facets, clear areas of (110) facets, and terraces on
the (111) facets.
\bigskip

Figure~2 shows the gradual melting of the cluster as the energy is
increased.  Figure~2a shows pre--melted (110) and (100) facets, with
the (110) regions appearing to form liquid lenses similar to those
predicted by L\"owen.\cite{Lowen}  At slightly higher energies, the
melted regions coalesce into a liquid envelope containing a crystalline
nucleus.  Figure~2b shows the crystalline nucleus near the energy at
which it becomes unstable.  We also did a simulation series starting
from a fully melted liquid cluster with a minuscule central fcc seed.
This formed a highly defective cluster with many stacking faults which
slightly modified the pre--melting behavior, but left the remainder of
the thermodynamics virtually unchanged.

\def\epsfsize#1#2{0.5#1}
{\bf Figure 2.}
Two cross sections of the
cluster, at energies (a) -3.065 and (b) -2.990 eV/atom.  The
cutting planes shown are (a) (100) and (b) (111).  The atomic
positions are averages over the equilibrated part of the
simulation.  The liquid atoms are shown in black, the
crystalline atoms in white.  A histogram of position
fluctuations was used to distinguish liquid from crystalline atoms.
\bigskip

The qualitative behavior of the dynamics in this simulation is also of
interest.  In the case of the highly defective cluster, we observed one
of the stacking faults relieve itself during the simulation, when a
partial dislocation nucleated and passed through the crystal.  The
dynamics of defect creation and annihilation will bear further study.
For both simulations series, the surface diffusion is rather rapid on
the melted surfaces, but moderately slow on the crystalline surfaces on
the time scales investigated.  Finally, we observe significant critical
slowing down of the equilibration between crystal and liquid near the
instability point.  Deep in the liquid and crystalline regions of
temperature, the equilibration time--scales were essentially zero; at
the energy shown in figure~2b, they slowed down to several tens of
picoseconds (requiring several days of Connection Machine time for
full equilibration).

In figure~3a, we plot $E(T)$, the total energy of the cluster versus
its temperature $T$.  For an infinite system at constant temperature,
this plot would consist of two straight lines and an abrupt vertical
jump at $T_c$ (ignoring the exponentially small, experimentally
unobservable effects of droplet fluctuations).  For the present
cluster, this simple jump unfolds.  The upper line represents the
liquid state.  Below around 1315~K the liquid becomes metastable, but
the nucleation barrier to form a crystalline nucleus remains
insurmountable until much lower temperatures.

The straight portion of the lower curve at low temperatures represents
the faceted crystal.  The slope of the curve starts to increase well
below $T_c$, representing the latent heat of pre--melting for the
various facets, starting with the open (110) surface.  There have been
experimental observations of superheating in metal
clusters,\cite{superheat} which could be due to pinning of the liquid
lens boundaries to the facet edges.\cite{Nozieres}  Superheating has
also been observed in simulations\cite{Manninen} for smaller magic--number
clusters, but it is likely that this superheating reflects the lack of
steps or adsorbed atoms on their facets.  We have observed no signs of
superheating or latent heat jumps here.  It is possible that the
pinning barriers are small in our cluster, but may become observable for
the much larger experimental clusters.

We find that the crystalline nucleus detaches
completely from the surface of the cluster below the temperature
maximum in the lower curve.  At higher energies, the temperature {\it
decreases} as the energy is increased (as observed in simulations
of  smaller clusters
by Labastie and Whetten\cite{Labastie}).  This is a simple consequence
of critical droplet theory.  The free energy per atom of a cluster with
a solid nucleus containing the fraction $\eta$ of the atoms in the
cluster is approximately $f(\eta)=-L [(T_c-T)/ T_c] \eta + \gamma
\eta^{2/3}$.  The first term represents the free energy difference
between the supercooled liquid and the crystal and $L$ is the latent
heat per atom.  The second term represents the solid-liquid interface
tension and the coefficient $\gamma$ is proportional to a suitable average
of the interfacial free energy density over the equilibrium crystal
shape. The forces from these two terms balance when $(T_c-T)/T_c =
(2/3) (\gamma / L) \eta^{-1/3}$ so the smaller the crystalline nucleus, the
larger the undercooling needed to stabilize it.  When energy is added
to the system, part of the crystal melts, absorbing the energy.  The
remainder of the crystal has a smaller radius of curvature, and the
interface tension demands a larger undercooling: even more of the
crystal melts leading to a net decrease in temperature.

\bigskip
\def\epsfsize#1#2{0.5#1}
{\bf Figure 3.}
In (a) is shown the $E(T)$
energy vs.~temperature plot for the copper cluster simulation
at constant energy.  The black points represent the energies
depicted in Fig.~2.  In (b) is shown $E(T)$ as derived from our
simple model, Eq.~\ref{eq:energy}.  Here, the lower solid curve
is the (partially) crystalline state, the upper solid curve is
the liquid state, the dashed curve represents the unstable
critical nucleus.  The dotted horizontal line denotes the
energy at which the entropies of the liquid and the mixed
phases are equal.
\bigskip

A rough estimate of the temperature range over which the transition is
broadened can be found by estimating the undercooling at which the
crystalline nucleus becomes unstable.  Equating the temperature drop
from melting a small region to the shift in the undercooling needed to
stabilize the now smaller crystalline region yields an instability
undercooling $\Delta T \sim (\gamma^3 / L^2 c N T_c)^{1/4} \sim
N^{-1/4}$, where here $c$ represents a weighted average of the specific
heats of the bulk liquid and solid.  This broadened transition is only
visible in constant--energy simulations; as $N \rightarrow \infty$, it
is larger than the finite--size effects at constant temperature
considered in previous work.\cite{Privman}  Using the more complete
analysis below, we have tested the size dependence of both the
crossover temperature (where the liquid entropy equals that of the
cluster) and the instability temperature (where the cluster becomes
absolutely unstable):  both scale as $N^{-1/4}$.

We can turn this simple explanation into a quantitative calculation.
First, because we work at constant energy, thermodynamics tells us that
we must maximize the entropy per atom $s$ (rather than minimize the
free energy $f$):
\begin{equation}
s(\eta,e_s,e_l) = \eta s_s(e_s) + (1-\eta) s_l(e_l).\label{eq:entropy}
\end{equation}
We here regard the entropy as a function of the solid fraction $\eta$,
the energy per atom in the solid nucleus $e_s$, and the energy per atom
in the liquid region $e_l$.  The entropy in the solid phase $s_s$ is
given as function of energy by $s_s(e_s)=s_s^c+c_s \log [
(e_s-e_s^c)/(T_c c_s) +1]$ where we make the approximation that the
specific heat $c_s$ is independent of temperature. The superscript $c$
indicates values at the transition point.  The entropy in the liquid
phase can be similarly defined, using that the energy difference per
atom between the two phases at the critical temperature is given by the
latent heat $L=e_l^c-e_s^c$, and the entropy difference is $s_l^c-s_s^c
= L/T_c$.

The energy per atom $e$ we write as
\begin{equation}
\FL e(\eta,e_s,e_l)=\eta e_s + (1-\eta) e_l
+ \frac{4 \pi}{N} [R^2 \gamma_{lv}
+ R_s^2(\gamma_{sl}+\Delta\gamma
\exp(-2(R-R_s) / \xi))],\label{eq:energy}
\end{equation}
with $\Delta\gamma = \gamma_{sv}-\gamma_{sl}-\gamma_{lv}$ where
$\gamma_{sv}$, $\gamma_{sl}$, and $\gamma_{lv}$ denote the free energy
densities of the solid--vapor, solid--liquid, and liquid--vapor
interfaces, respectively. The radius of the cluster $R$ and the radius
of the solid nucleus $R_s$ can be expressed in terms of the densities
and the solid fraction $\eta$.  The last term in Eq.~\ref{eq:energy}
represents the interaction between the solid--liquid and the
liquid--vapor interfaces and gives rise to premelting of a surface if
$\Delta\gamma$ is positive.  As usual,\cite{Landau} the ambiguity of
where to place the liquid--solid interfacial position is reflected in
the breakup of the surface free energy into entropy and energy:  we use
the convention that attributes the free energy cost entirely to
energy.

The experimental values of most of the parameters entering
Eqs.~\ref{eq:entropy} and \ref{eq:energy} are known.\cite{CRC,Pluis}
There is substantial uncertainty in only two:  The solid-vapor and
solid-liquid interfacial energies.  The value for the solid--liquid
interfacial energy $\gamma_{sl}$ we take as\cite{Pluis} 263 mJ/m$^2$.
However, values as low as 177 mJ/m$^2$ have been found.\cite{Turnbull}
The value for the solid--vapor interfacial energy $\gamma_{sv}$ is
poorly known.  Here, we take the value from ref.~\cite{Pluis}.  A
change in this parameter of only 1 percent will change the
$\Delta\gamma$ by a factor of 2.  This parameter controls the onset of
pre--melting, and hence the shape of $E(T)$.  With the present model
(ignoring anisotropy as well as faceting and disorder on edges and
vertices) and the given experimental values, we do not expect to
reproduce in detail the pre--melting observed in the simulations.  The
exponential decay of the interaction between the solid--liquid and
liquid--vapor interfaces follows from a Ginzburg--Landau
analysis\cite{Pluis} of pre--melting of flat surfaces.  The correlation
length $\xi$ can be estimated from the solid--liquid interfacial free
energy density using the Hansen--Verlet melting rule.\cite{Pluis}
Maximizing the entropy, Eq.~\ref{eq:entropy}, with respect to $e_s$ and
$e_l$ at fixed energy $e$, Eq.~\ref{eq:energy}, gives rise to the
natural condition that the solid and liquid parts of the cluster must
have the same temperature, and a further maximization with respect to
the solid fraction $\eta$ leads to the $E(T)$ curves shown in
figure~3b.  At constant temperature, the transitions are vertical on
this plot:  there are two metastable states separated by the critical
nucleus.  The pre--melted crystal becomes unstable when the $E(T)$
curve has a vertical tangent.  At constant energy, on the other hand,
the crystalline nucleus surrounded by liquid becomes unstable when the
tangent is horizontal.

Comparing figures~3a and 3b, we find that the molecular dynamics
simulations using effective--medium theory potentials are in
surprisingly good agreement with our simple model that uses only
experimental data for the specific heats, the latent heat, and the
effects of surface tension.  However, there is a fairly substantial
temperature shift between the figures.  The experimental transition
temperature for bulk copper is 1356.2~K: the interfacial tensions
depress the transition in the model cluster to about 1190~K as seen in
figure~3b.  On the other hand, the transition is seen at 1335~K in the
simulations.  This could in part be due to the large uncertainty in the
experimental value for the crystal--liquid surface tension used in our
simple model, but it is also conceivable that the effective--medium
theory\cite{Norskov,Stoltze} transition temperature could be off by the
necessary 145~K.  Simulations for a range of cluster sizes could
pinpoint these properties.

We have shown that simple critical droplet theory, developed to study
nucleation rates, provides a complete explanation for our copper
cluster melting problem as studied through molecular dynamics
simulations.  We conclude by using the theory to calculate the
nucleation rate for constant total energy.  The dotted horizontal line
in figure~3b shows the energy $e_{\rm cross}$ at which the entropy of
the liquid state equals that of the mixed state, which at this point
has about 7200 atoms in the crystalline nucleus.  Below $e_{\rm cross}$
the liquid droplet is supercooled and metastable, as is the mixed phase
above this energy.  Near $e_{\rm cross}$, the system is in principle in
a mixed state, where the crystalline nucleus appears and disappears
with thermal fluctuations.  The {\it time} needed to see a fluctuation,
though, is huge.  The intersection of the dotted line with the middle,
dashed curve represents a transition state with a smaller, crystalline
nucleus of the same energy and consisting of approximately 1100 atoms.
However, this critical nucleus has an entropy lower by about $170 k_B$,
which is thus the entropy barrier between the two metastable states.
The nucleation rate of the crystal from the supercooled liquid is
consequently some microscopic prefactor times $e^{-170}$.  We conclude
that the nucleation rate is negligible both in simulations and
experimentally until temperatures much lower than $T_c$.

\bigskip
\acknowledgments

Financial support from the Center for Surface Reactivity
is gratefully acknowledged.
The Connection Machine resources at UNI$\bullet$C were provided through
the Center for Parallel Computer Research (CAP) as funded by the Danish
Natural Science Research Council and the Danish Technical Research
Council.
JPS would like to acknowledge DOE grant \#DE-FG02-88-ER45364
for support, and would like to thank the Technical University of
Denmark and NORDITA for support and hospitality.

\end{document}